\documentclass[aps,prb,twocolumn,superscriptaddress,showpacs,longbibliography,float]{revtex4-1}
\usepackage{graphicx,bm,mathtools,pbox}
\usepackage[pdfstartview=FitH, CJKbookmarks=true, bookmarksnumbered=true, bookmarksopen=true, colorlinks, pdfborder=001, linkcolor=blue, anchorcolor=blue, citecolor=blue]{hyperref}

\begin{document}

\title{Magnetic order in  Nd$_2$PdSi$_3$ investigated using neutron scattering and muon spin relaxation}

\author{M. Smidman}
\email[]{msmidman@zju.edu.cn}
\affiliation{Center for Correlated Matter and Department of Physics, Zhejiang University, Hangzhou 310058, China}
\author{C. Ritter}
\affiliation{Institut Laue Langevin, BP 156, 38042 Grenoble Cedex 9, France}
\author{D. T. Adroja}
\email[]{devashibhai.adroja@stfc.ac.uk}
\affiliation{ISIS Facility, STFC, Rutherford Appleton Laboratory, Chilton, Didcot, Oxfordshire OX11 0QX, United Kingdom}
\affiliation{Highly Correlated Matter Research Group, Physics Department, University of Johannesburg, PO Box 524, Auckland Park 2006, South Africa}
\author{S. Rayaprol}
\affiliation{UGC-DAE Consortium for Scientific Research, Mumbai Centre, R-5 Shed, BARC Campus, Trombay, Mumbai – 400085, India}
\author{T. Basu}
\affiliation{Tata Institute of Fundamental Research, Homi Bhabha Road, Colaba, Mumbai 400005, India}
\author{E. V. Sampathkumaran}
\affiliation{Tata Institute of Fundamental Research, Homi Bhabha Road, Colaba, Mumbai 400005, India}
\author{A. D. Hillier}
\affiliation{ISIS Facility, STFC, Rutherford Appleton Laboratory, Chilton, Didcot, Oxfordshire OX11 0QX, United Kingdom}
\date{\today}

\begin{abstract}
The rare-earth based ternary intermetallic  compounds $R_2TX_3$ ($R$ = rare-earth, $T$ = transition-metal, $X$ = Si, Ge, Ga, In) have attracted considerable interest   due to a wide range of interesting low temperature properties. Here we investigate the magnetic  state of Nd$_{2}$PdSi$_{3}$ using neutron diffraction, muon spin relaxation ($\mu$SR) and  inelastic neutron scattering (INS). This compound appears anomalous among the $R_{2}$PdSi$_{3}$  series, since it was proposed to order  ferromagnetically, whereas others in this series are antiferromagnets. Although some members of the $R_2TX_3$ series have been reported to form ordered superstructures, our data are well described by Nd$_{2}$PdSi$_{3}$ adopting the AlB$_2$-type structure with a single Nd site, and we do not find evidence for superlattice peaks in neutron diffraction. Our results confirm the onset of long range magnetic order below  $T_0=17$~K, where the whole sample  enters the ordered state. Neutron diffraction measurements establish the presence of a ferromagnetic component in this compound, as well as  an antiferromagnetic one which has a propagation vector $\mathbf{k_2}=(1/2,1/2,1/4-\delta)$ with a temperature dependent $\delta\approx0.02-0.04$, and moments orientated exclusively along the $c$-axis.  $\mu$SR measurements suggest that these components coexist on a microscopic level, and therefore the  magnetic structure of Nd$_{2}$PdSi$_{3}$ is predominantly ferromagnetic, with a sinusoidally modulated antiferromagnetic contribution which reaches a maximum amplitude at 11~K, and becomes smaller upon further decreasing the temperature. INS results show the presence of crystalline-electric field (CEF) excitations above  $T_0$, and from our analysis we propose a CEF level scheme.
\end{abstract}

\maketitle

\section{Introduction}

Intermetallic compounds containing rare earth ($R$) or actinide atoms with partially filled $4f$ or $5f$ shells allow for the realization of complex magnetic ground states, which can be driven by the presence of competing magnetic interactions ~\cite{jensen1991rare}.  In 
addition to the intersite Ruderman-Kittel-Kasuya-Yosida (RKKY) interaction which gives rise to a magnetically ordered ground state,  the localized $f$-electrons can also hybridize with the conduction electrons,  resulting in an onsite Kondo effect which competes with the RKKY interaction \cite{SDoniach1977}. As the strength of the Kondo interaction is increased, the magnitude of the ordered magnetic moment is reduced, eventually yielding a  non-magnetic ground state. In many systems there is a critical value of the interaction strength at which the magnetic ordering temperature is suppressed to zero temperature. In the vicinity of this quantum critical point (QCP), quantum fluctuations dominate the physical properties instead of classical thermal fluctuations ~\cite{AJHertz1976, AJMillis1993, TMoriay1995,TMoriay2000}. Here there is a breakdown of Landau Fermi-Liquid theory, and the system exhibits non-Fermi-liquid behaviour ~\cite{GRStewart2001,PColeman2015}. Moreover, some Ce- and Yb-based systems exhibit unconventional heavy fermion superconductivity near an antiferromagnetic QCP, where some of the most prominent examples crystallize in the tetragonal ThCr$_2$Si$_2$-structure (1-2-2 family) \cite{CeCu2Si2,YbRh2Si2,CePd2Si2}. As such, rare earth intermetallics have been fertile grounds for novel solid state phenomena, in particular those driven by competing electronic interactions.

In this respect, the ternary rare earth containing compounds $R_2TX_3$ ($R$=rare-earth, $T$=transition metals and $X=$Si, Ge, In, Ga) have attracted considerable attention. Just as in the 1-2-2 families,  wide-ranging anomalies and phenomena have been reported (even in Gd-based systems),  including competition between the Kondo effect and magnetic ordering, re-entrant spin-glass phases, geometrically frustrated and low-dimensional magnetism,  non-Fermi liquid behavior,    giant magnetoresistance, and an anisotropic magnetocaloric effect \cite{Das1994,Mallik1998a,Mallik1998b,Majumdar1999,Saha1999,Saha2000,Sampath2000,Majumdar2000a,Majumdar2000b,Majumdar2001,Gondek2002,Majumdar2002,Paulose2003,Nakano2007,Iyer2007,Swapnil2008,Swapnil2010,Xu2011,KMukherjee2011,R2PdSi32,DLLi2003,R2PdGe31}. One reason for the rich range of properties  is that there are several structural variations with this composition, depending on the nature of $T$ and $X$ ions, including hexagonal  (space groups $P6/mmm$, $P\bar{6}2c$, $P6_3/mmc$) and orthorhombic ($Fmmm$) materials \cite{Chevalier1984,R2PdSi36,Gladyshevskii1992,Gordon1997,Tang,R2PdSi32}. In particular, a number of these consist of  honeycomb  layers of $T$ and $X$ separated by triangularly arranged $R$ ions \cite{R2PdSi36,Das1994}, which can give rise to  frustration in the presence of nearest neighbor antiferromagnetic interactions.

In this article, we focus on the Nd compound in the $R_2$PdSi$_3$ series, which  is an unusual case, since Nd$_2$PdSi$_3$ is believed to undergo a ferromagnetic transition near 16~K \cite{KMukherjee2011,R2PdSi32,DLLi2003,Xu2011}, while the materials with other $R$ ions are antiferromagnetic at the onset of magnetic order \cite{R2PdSi36}. In addition, there is a strong enhancement of  the magnetic ordering temperature ($T_0$)  compared to that expected from de Gennes scaling in this compound.  Mukherjee \textit{et al} \cite{KMukherjee2011} concluded that $4f$ hybridization effects, typical of Ce systems but uncommon among Nd systems, play a role in the anomalous magnetism of this material. It is also important to note that while the strong $\lambda$-anomaly in the heat-capacity  at $T_0$ and  increase of the transition temperature with increasing  magnetic field are consistent with the onset of long range ferromagnetic order near 16~K \cite{KMukherjee2011},  there is also a small frequency dispersion in the ac magnetic susceptibility at $T_0$, which is typical of spin-glasses as though antiferromagnetism competes at the transition \cite{KMukherjee2011,DLLi2003}. 

This situation warrants further detailed investigations using microscopic methods to understand the unusual magnetic properties. As a result, we carried out  powder neutron diffraction, muon-spin relaxation ($\mu$SR) and inelastic neutron scattering (INS) measurements on Nd$_2$PdSi$_3$. Our neutron diffraction study clearly reveals  ferromagnetic and antiferromagnetic Bragg peaks below the ordering temperature,   while the $\mu$SR study shows that this transition corresponds to the onset of bulk long range magnetic order with a lack of macroscopic phase separation between the antiferromagnetic and ferromagnetic components. Inelastic neutron scattering shows the presence of crystalline electric field (CEF) excitations above $T_0$, which can be accounted for by a CEF model for the splitting of the $J=9/2$ multiplet of Nd$^{3+}$. Meanwhile an additional excitation below $T_0$ is observed, which likely corresponds to the Zeeman splitting of the Kramer's doublets by a molecular field.

\begin{figure}[t]
\begin{center}
  \includegraphics[width=0.7\columnwidth]{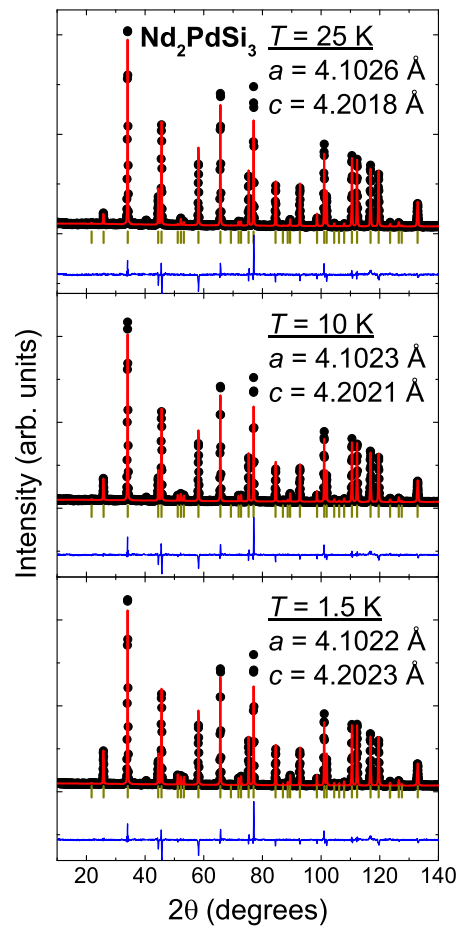}
\end{center}
	\caption{(Color online) High resolution  powder neutron diffraction measurements of Nd$_2$PdSi$_3$ at three temperatures, measured on the D2B instrument. At 25~K, the solid red line shows the refinement of the crystal structure, while the ticks and blue lines below the data show the peak positions and difference plot respectively. At 1.5~K and 10~K, the data were refined taking into account the crystal structure and a ferromagnetic component, where the scale factor for the nuclear component was fixed from the 25~K refinement. Here the ticks correspond to the position of both the structural Bragg peaks, as well as the magnetic Bragg peaks corresponding to the ferromagnetic component.}
   \label{Fig1}
\end{figure}
\section{Experimental details}

Polycrystalline samples of Nd$_2$PdSi$_3$ were prepared by arc-melting the constituent elements in a stoichiometric ratio. Inelastic neutron scattering  and $\mu$SR measurements were performed at  the ISIS facility at the Rutherford Appleton Laboratory, U.K. INS measurements were performed  using the high neutron flux MERLIN time-of-flight (TOF) spectrometer, with incident energies of 15 and 45~meV (elastic resolutions of 1.3 and 2.9~meV)  selected using a Fermi chopper. $\mu$SR measurements were carried out using the EMU spectrometer. Powder neutron diffraction experiments were performed using the high resolution D2B and the high intensity D20 powder diffractometers at the Institut Laue Langevin (ILL), Grenoble, France. The high resolution data were taken at $T$~=~1.5, 10.5 and 25~K while on D20 a temperature dependent ramp was measured between 1.7~K and 25~K with longer acquisition times at $T$~=~1.7~K, 11~K and 25~K.

\begin{figure}[!h]
\begin{center}
  \includegraphics[width=0.8\columnwidth]{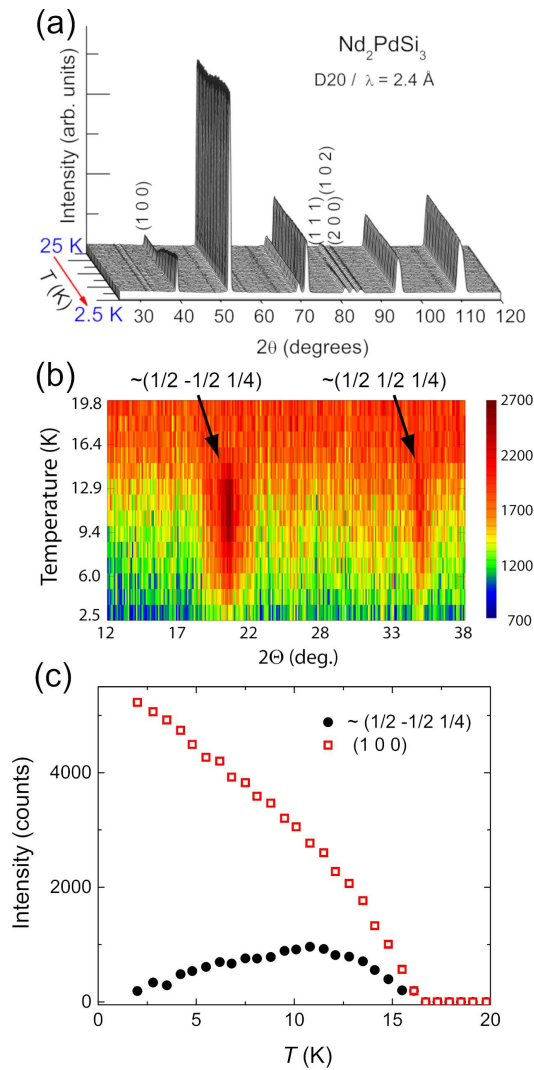}
\end{center}
	\caption{(Color online) (a) Diffraction patterns of  Nd$_2$PdSi$_3$ measured using the D20 instrument at various temperatures below 25~K. At the magnetic transition there is a clear increase in intensity of some of the nuclear peaks, which corresponds to the onset of an ordered ferromagnetic component. In addition, there is the emergence of peaks away from the nuclear peaks, corresponding to an antiferromagnetic component with a propagation vector of about $\mathbf{k}\approx$(1/2,1/2,1/4).  (b) Thermodiffractogram across a limited angular range for temperatures between 2.5 and 20~K, where the color plot represents the intensity. The emergence of two antiferromagnetic Bragg reflections below the ordering temperature can clearly be resolved. (c) Magnetic contributions to the $(1~0~0$) and $(1/2~-1/2~1/4 )$ Bragg reflections as a function of temperature after subtracting the high temperature data, corresponding to the ferromagnetic and antiferromagnetic components respectively. Note that the magnetic contribution to both peaks onset at the same temperature and while the intensity of the (1~0~0) peak continues to increase with decreasing temperature, the intensity of $(1/2~-1/2~1/4 )$ reaches a maximum at around 11~K.}
   \label{Fig2}
\end{figure}

\section{Results and discussion}

\subsection{Neutron diffraction}

Figure~\ref{Fig1} displays the neutron  powder diffraction patterns of Nd$_2$PdSi$_3$ at three temperatures, measured using the D2B diffractometer with a neutron wavelength of 1.59~\AA. The data at 25~K was collected at a higher temperature than $T_0$, and therefore the Bragg peaks entirely originate from the crystal structure. Due to the high resolution of the D2B instrument, these data are particularly suitable for refining the crystal structure. The crystal structure refinement was performed using the FullProf software \cite{Carvajal1993}, and the results are displayed in Fig.~\ref{Fig1}(a). The lattice parameters at 25~K of $a=4.1026(1)$~\AA~ and $c=4.2018(1)$~\AA~ are consistent with previous results \cite{R2PdSi32}. The data were refined using the AlB$_2$-type structure, where both Pd and Si occupy a single crystallographic site, and there is no indication in these powder data that these atoms form a superstructure. We note that while such superstructure peaks may be very weak and therefore difficult to detect, in a previous study it was reported that the AlB$_2$-type structure accounts best for the data \cite{R2PdSi32}.

Some additional intensity on some of the nuclear Bragg reflections is visible at 1.5~K and 10~K, indicating the presence of magnetic scattering with a propagation vector $\mathbf{k_1}=0$. Magnetic symmetry analysis using the program BASIREPS \cite{Carvajal,Ritter2011} for the Wyckoff site 1a of Nd in $P6/mmm$ indicated the presence of two allowed irreducible representations (IRREP). One of these has a single basis vector (BV) pointing in the direction of the hexagonal $c$-axis and the second one has two BVs along the $a$- and $b$-axes of the hexagonal basal plane. Only the alignment of the ferromagnetic moments along the $c$-axis allowed the refinement of the data. The results are displayed in Figs.~\ref{Fig1}(b) and (c), where the values of the Nd moments ($\mu_{\rm Nd}$) are 1.50(4)$\mu_B$/Nd at 10~K and 2.08(4)$\mu_B$/Nd at 1.5~K.

Previous results  had suggested the presence of an antiferromagnetic  component to the magnetism \cite{KMukherjee2011,DLLi2003}, however, no sign of additional magnetic peaks could be detected in the data recorded on the D2B instrument displayed in Fig.~\ref{Fig1}. Therefore measurements were performed using the D20 diffractometer, which is more suited for resolving low intensity magnetic peaks due to a much higher neutron flux. The measurements were performed with a neutron wavelength of 2.4\AA, and therefore magnetic Bragg peaks at smaller momentum transfers are also more readily accessible. The diffraction patterns measured on D20 at a large number of temperatures below 25~K are displayed in Fig.~\ref{Fig2}(a). Apart from some nuclear peaks showing a significant increase in intensity at the ordering temperature, such as the (1~0~0) reflection, which are linked to the aforementioned ferromagnetic order, as shown in Fig.~\ref{Fig2}(b), new peaks are found to emerge at the same temperature. These new, additional peaks are found at positions which do not coincide with the nuclear reflections, but are consistent with an antiferromagnetic component with a propagation vector of about $\mathbf{k_2}\approx(1/2,1/2,1/4)$. Figure~\ref{Fig2}(c) displays the magnetic contribution to the peak intensity for two reflections, namely the $(1~0~0)$ reflection corresponding to the ferromagnetic component, and the new  $(1/2~-1/2~1/4)$ reflection, which has the largest intensity of the Bragg peaks corresponding to the additional antiferromagnetic order. We note that in the data measured using the D2B instrument, this antiferromagnetic peak would be expected at a low scattering  angle where there is a large angle dependent background, and therefore this cannot be detected.  It can be seen that both components set in at the same temperature of around 16~K. However the magnetic intensity from the ferromagnetic reflection increases monotonically with decreasing temperature, while the significantly smaller antiferromagnetic component reaches a maximum intensity at around 11~K, before decreasing and nearly disappearing again  upon approaching the lowest temperatures.

\begin{figure}[tb]
\begin{center}
  \includegraphics[width=0.95\columnwidth]{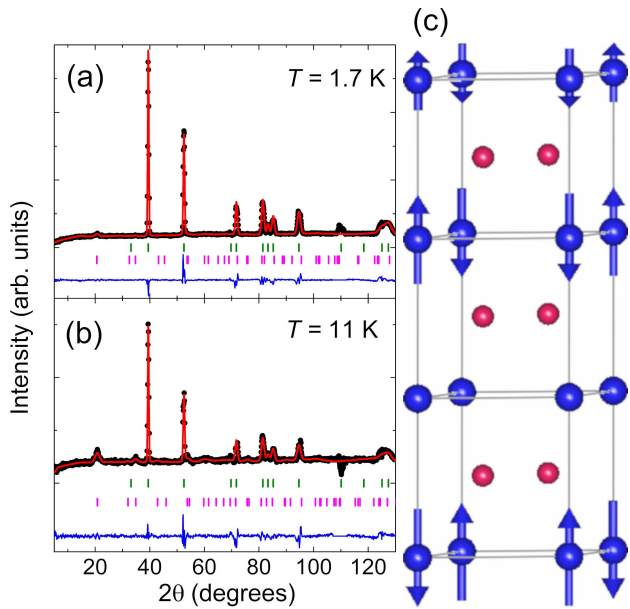}
\end{center}
	\caption{(Color online) (a) Diffraction patterns of  Nd$_2$PdSi$_3$ measured with the D20 instrument after subtracting high temperature (25~K) data  at (a) 1.7~K, and (b) 11~K. The solid red lines show refinements to a magnetic structure with both a ferromagnetic component and an antiferromagnetic component with  a propagation vector of $\mathbf{k_2}=(1/2,1/2,1/4-\delta)$ ($\delta\approx0.02-0.04$). The ticks show the theoretical positions of the peaks corresponding to the two components, while the blue line shows the difference plot. (c) The crystal structure of Nd$_2$PdSi$_3$ along with the magnetic structure corresponding to the antiferromagnetic component, with Nd atoms in blue, and Pd/Si atoms in red. Note that for some Nd layers, the antiferromagnetic component is either very small or zero.}
   \label{Fig3}
\end{figure}

The purely magnetic contributions to the neutron diffraction data taken with better statistics at 1.7~K and 11~K are displayed in Figs.~\ref{Fig3}(a) and (b), respectively, which were obtained by subtracting the 25~K data. A magnetic symmetry analysis for $\mathbf{k_2}=(1/2,1/2,1/4)$ resulted in three allowed IRREPs, each having one BV. Only the IRREP having its BV along the hexagonal $c$-axis is able to correctly describe the measured intensities of the antiferromagnetic peaks. The refinement of the purely magnetic data was performed by fixing the scale factor to the value determined from the refinement of the purely nuclear 25~K data \cite{RefNote1}. The presence of two magnetic propagation vectors describing the ferromagnetic and antiferromagnetic components can be interpreted as either reflecting a phase separation scenario, where one phase adopts a ferromagnetic  structure while the second one is antiferromagnetic, or as a single phase with both magnetic couplings embracing the whole sample volume. It is not possible to discriminate between these two scenarios from the refinements, as long as only a single nuclear phase is resolved. Both options were therefore tested, where in the phase separation picture the magnetic moment was constrained to have the same value in the antiferromagnetic and ferromagnetic phases, and the phase fractions were determined  by splitting the fixed scale factor.  In the second model where both components coexist throughout the volume of the sample, the ferromagnetic moments are refined to 1.97(1)$\mu_B$/Nd at 1.7~K and 1.36(1)$\mu_B$/Nd at 11~K, while the antiferromagnetic component has respective moments of 0.58(4)$\mu_B$/Nd and 1.00(2)$\mu_B$/Nd (Fig.~\ref{Fig3}). We note that the presence of an antiferromagnetic component in this data is not inconsistent with the data measured on D2B, since given the magnitude of these antiferromagnetic moments, together with the longer neutron wavelength, the antiferromagnetic  Bragg peaks would not be expected to observable on D2B above the uncertainty of the background.

During these refinements it became apparent that the exact value of the $l$-component of the antiferromagnetic propagation vector $\mathbf{k_2}=(h,k,l)$ assumes an incommensurate value of $l$~=~0.234(4) at 11~K. Figure~\ref{Fig3}(c) displays the magnetic structure of this antiferromagnetic component. This corresponds to a sinusoidal modulation of the spins running along the $c$-axis with $\mu_{\rm Nd}$ of the refinement being the amplitude of the modulation, while in the $a$ and $b$ directions adjacent spins are antialigned. Similar magnetic propagation vectors were found from neutron diffraction measurements of Ce$_2$PdSi$_3$ \cite{R2PdSi32}, however in this case the moments are ferromagnetically aligned in the $ab$-plane, perpendicular  to the modulation direction. On the other hand, the magnetic structures of $R_2$PdSi$_3$ for $R=$Tb, Dy, Er and Ho are quite different, where the propagation vector has an incommensurate modulation within the hexagonal plane. Note that the propagation vectors (1/2,1/2,1/4) and (1/2,1/2,0.234) are from the point of view of magnetic symmetry identical. In fact a value of $l$~=~0.25 would see the nodal points of the modulation being positioned exactly on every fourth atom along the $c$-axis. The fit of the data at 1.7~K led to a refined value of $l$~=~0.215(8), indicating a small change of the modulation wavelength. Refining the data assuming the phase separation between antiferromagnetic and ferromagnetic regions yields at $T$~=~11~K, 62(1)\% of the sample being ferromagnetic and 38(1)\% being antiferromagnetic with $\mu_{\rm Nd}=1.71(1)\mu_B$/Nd, while at 1.7~K the respective fractions are 92(1)\% and 8(1)\% with  $\mu_{\rm Nd}=2.06(2)\mu_B$/Nd.

\subsection{Zero-Field $\mu$SR measurements}

\begin{figure}[t]
\begin{center}
  \includegraphics[width=0.99\columnwidth]{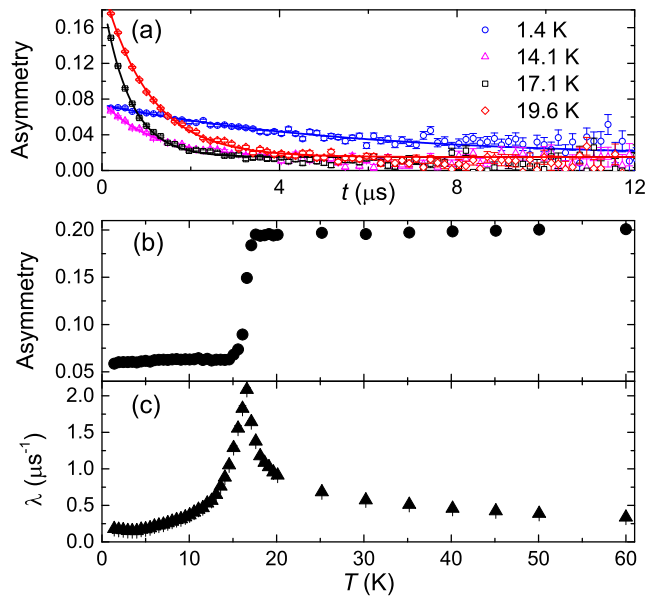}
\end{center}
	\caption{(Color online) (a) Zero field $\mu$SR spectra of Nd$_2$PdSi$_3$ measured at four temperatures, both above and below the magnetic transition. The solid lines correspond to the fitting described in the text, where (b) shows the initial asymmetry of the component corresponding to the sample $A_1$, while (c) shows the Lorentzian relaxation rate $\lambda$.  }
   \label{Fig4}
\end{figure}

Zero-field muon spin relaxation ($\mu$SR) measurements at selected temperatures are displayed in Fig.~\ref{Fig4}(a). It can be seen that at lower temperatures there is a significant drop in the asymmetry, consistent with the onset of long range magnetic order. We do not observe a clear signature of coherent oscillations in the time dependent asymmetry spectra below $T_0$, which indicates that the size of the ordered Nd moment is such  that the local field at the muon stopping site is too large for the corresponding oscillations to be resolved, due to the finite width of the ISIS muon pulse. The asymmetry spectra were fitted with an exponential decay, suitable for rapidly fluctuating moments,

\begin{equation}
A(t) = A_0 + A_1{\rm e}^{-\lambda t}
\end{equation}

\noindent where $A_0$ is the background term arising from the muons stopping on the silver sample holder, $A_1$ is the initial asymmetry corresponding to muons stopped in the sample, and $\lambda$ is the Lorentzian relaxation rate. The value of $A_0$ was fixed from the analysis at 60~K, while the temperature dependence of the fitted  $A_1$  is displayed in Fig.~\ref{Fig4}(b). It can be seen that  there is a sharp drop in $A_1$ setting in at around 17~K, below which there is a decrease to a value of around one third of that at high temperatures. Upon further lowering the temperature, there is little change in $A_1$. Such a loss of  asymmetry is expected from $\mu$SR measurements of polycrystalline magnetically ordered materials, where the asymmetry corresponding to two thirds of the implanted muons depolarizes more rapidly than the time frame of the $\mu$SR measurements. This strongly suggests that the whole sample magnetically orders at the magnetic transition at $T_0$=17~K. As noted in the previous section, from refinements of the neutron diffraction data, we cannot distinguish between the scenarios of phase separated antiferromagnetic and ferromagnetic regions, and coexistence of the two phases. The fact that the $\mu$SR spectra can be well fitted with only one component despite the presence of both antiferromagnetic and ferromagnetic Bragg peaks, as well as these  onsetting at a single transition, suggests that there is indeed microscopic coexistence of these ordered components and no phase separation. The relaxation rate $\lambda$ (Fig.~\ref{Fig4}(c)) shows a sharp peak around the magnetic transition, which is consistent with the presence of spin fluctuations which critically slow down upon approaching the transition at 17~K.

\begin{figure}[t]
\begin{center}
  \includegraphics[width=0.99\columnwidth]{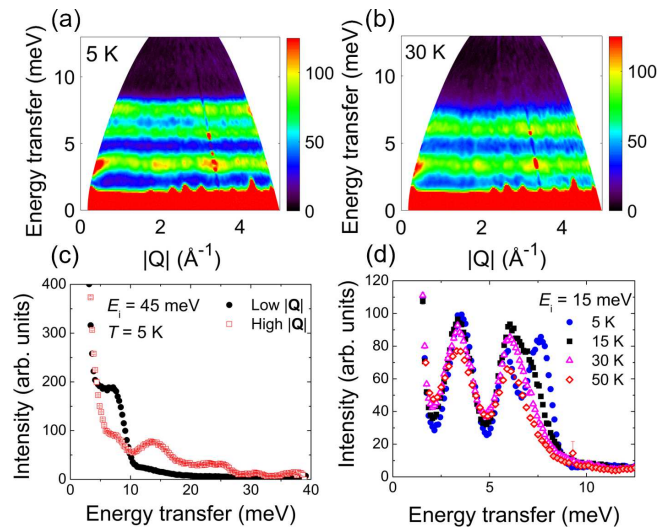}
\end{center}
	\caption{(Color online)  Color-coded plot of the inelastic neutron scattering intensity as a function of energy transfer and momentum transfer, at (a) 5~K, and (b) 30~K, with an incident neutron energy of $E_i=15$~meV. The scattering intensity scale is in arbitrary units. (c) Cuts of the neutron scattering intensity as a function of energy transfer obtained from integrating across a low $|\mathbf{Q}|$ range ($0-3$\AA$^{-1}$) and a high $|\mathbf{Q}|$ range ($6-9$\AA$^{-1}$)  for  $E_i=45$~meV at 5~K. (d) Cuts of the neutron scattering intensity as a function of energy transfer obtained from integrating across a low $|\mathbf{Q}|$ range ($0-3$\AA$^{-1}$)  at various temperatures for  $E_i=15$~meV.}
   \label{Fig5}
\end{figure}

\subsection{Inelastic neutron scattering measurements}

\begin{figure}[t]
\begin{center}
  \includegraphics[width=0.9\columnwidth]{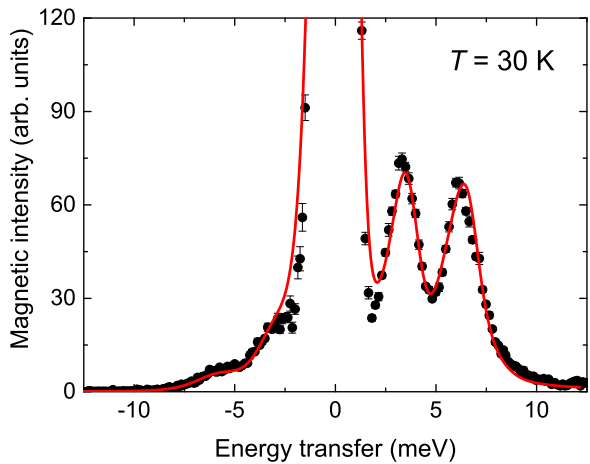}
\end{center}
	\caption{(Color online) Magnetic contribution to the inelastic neutron scattering intensity at 30~K, estimated from subtracting the scaled high $|\mathbf{Q}|$ data ($3-5$\AA$^{-1}$), from low $|\mathbf{Q}|$ data ($0-2$\AA$^{-1}$). The solid lines show a fit to the CEF model described in the text.  }
   \label{Fig6}
\end{figure}

Inelastic neutron scattering measurements were also performed using the MERLIN spectrometer, with incident energies ($E_i$) of 15 and 45~meV, to investigate the CEF excitations ($T>T_0$) as well as spin wave excitations below the ordering temperature. The 2D color plots of the scattering intensity for energy transfer vs momentum transfer,  at 5~K and 30~K  are  displayed for $E_i$=15~meV in Figs.~\ref{Fig5}(a) and \ref{Fig5}(b), respectively  Three clear low energy magnetic excitations are observed at energies of around 3.6, 6, and 7.7~meV at 5~K (below $T_0$). Meanwhile above the ordering temperature (at  $T$=30~K), the high energy excitation near 7.7 meV is no longer observed, as shown for the color plot for the measurements at 30~K displayed in  Fig.~\ref{Fig5}(b). The measurements performed at $E_i=45$~meV do not reveal any evidence for magnetic excitations at higher incident energies [Fig.~\ref{Fig5}(c)]. This can be clearly seen from a comparison of cuts integrated over low momentum transfers $|\textbf{Q}|$ (0-3\AA$^{-1}$) and high $|\textbf{Q}|$ (6-9\AA$^{-1}$). Here while several peaks are found in the high  $|\textbf{Q}|$  cut, which likely correspond to phonon excitations,  no features are observed for the low $|\textbf{Q}|$ cut.

Low $|\textbf{Q}|$ cuts for $E_{\rm i}=15$~meV showing the magnetic excitations at four temperatures are shown in Fig.~\ref{Fig5}(d). The 3.6 and 6~meV excitations are found above $T_0$, and therefore these  likely correspond to CEF excitations. On the other hand, the  7.7~meV excitation at 5~K shifts to lower energies at 15~K, before disappearing in the paramagnetic state, suggesting that this corresponds to a spin-wave excitation. This excitation existing only  below  $T_0$ may also be interpreted as  arising due to Zeeman splitting of the CEF excitations in the presence of  a molecular field from the ordered Nd moments.  For a Nd$^{3+}$ ion with a 4f$^3$ electronic configuration  in a hexagonal CEF, the $J=9/2$ ground state multiplet is expected to split into five Kramer's doublets in the paramagnetic state. The corresponding Hamiltonian for Nd$^{3+}$ with $D_{6h}$ point symmetry is given by

\begin{equation}
\mathcal{H}_{\rm{CF}} = B_2^0{\rm{O_2^0}} + B_4^0{\rm{O_4^0}} + B_6^0{\rm{O_6^0}}+ B_6^6{\rm{O_6^6}}
\end{equation}

\noindent where $B_n^m$ are parameters and $\rm{O_n^m}$ are the Steven's operator equivalents.  The data at 30~K were analyzed using the above Hamiltonian, and the results are displayed in Fig.~\ref{Fig6}. The obtained  parameters are $B_2^0=-0.0843$~meV$, B_4^0=0.00244$~meV, $B_6^0=-5.49\times10^{-5}$~meV, and $B_6^6=1.04\times10^{-4}$~meV. This gives rise to five doublets, where the energy differences  from the ground state to the four excited levels are 3.44, 5.50, 6.45, and 13.75~meV. Here the wave function of the ground state doublet corresponds to $|\psi^{\pm} \rangle = 0.988\left|\pm \frac{5}{2} \right \rangle - 0.155\left|\mp \frac{7}{2} \right \rangle$. From this the ground state magnetic moments along the $c$-axis and in the $ab$-plane are 
$\mu_z=1.71\mu_{\rm B}$/Nd and $\mu_x=0.46\mu_{\rm B}/$Nd, respectively. If the anisotropy energy is calculated using the fitted parameters \cite{Marusi1990}, the easy axis is predicted to be along the $c$-axis, which is in agreement with  that observed in the single crystal susceptibility \cite{Xu2011}, and is the direction of the ordered moment below $T_0$. The fitted linewidth of the inelastic peaks corresponding to CEF excitations is 1.65~meV, which is slightly broader than the instrument resolution (the resolution at 5.9~meV is about 0.7~meV). This broadening may be a reflection of disorder between Pd and Si, since these occupy the same crystallographic sites in the AlB$_2$-type structure. In addition, linewidth broadening of CEF excitations may arise due to hybridization between Nd $4f$- and conduction electrons, as proposed from photoemission spectroscopy \cite{Maiti2019}.

Note that if Nd$_2$PdSi$_3$ were to adopt the crystal structure with the ordered superstructure (space group $P6_3/mmc$), then there would be  two inequivalent Nd crystallographic sites in the crystal structure. Since this would lead to different local environments for each Nd site, and hence different crystal field potentials, this would be expected to give rise to additional excitations from the ground state to excited CEF levels. The lack of additional excitations therefore is evidence for there being only one site for Nd, which further supports the crystal structure of Nd$_2$PdSi$_3$ corresponding to that displayed in Fig~\ref{Fig3}(c) with space group $P6/mmm$.

\section{Conclusions}
We have addressed the magnetic behavior of the anomalous Nd-based compound  Nd$_{2}$PdSi$_{3}$ using neutron diffraction, $\mu$SR measurements and inelastic neutron scattering. Neutron diffraction results reveals the presence of long range magnetic order, where magnetic Bragg peaks corresponding to both ferromagnetic and antiferromagnetic components setting in below $T_0$=17~K, where the latter correspond to a propagation wavevector $\mathbf{k_2}=(1/2,1/2,1/4-\delta)$ ($\delta\approx0.02-0.04$).  Moreover, the intensity of the ferromagnetic peaks continues to increase with decreasing temperature, while the intensity for the antiferromagnetic peaks reaches a maximum at around 11~K. A refinement of the magnetic structure reveals that the antiferromagnetic structure consists of  a sinusoidally modulated arrangement of spins along the $c$-axis, which are  antiferromagnetically coupled along the $a$ and $b$ directions, with the moments parallel to the modulation direction. The  $\mu$SR measurements further confirmed the onset of long range magnetic order, where the whole sample appears to undergo a magnetic transition below $T_0$, where the spectra can be fitted with a single relaxing component. This suggests the microscopic coexistence of the ferromagnetism  and antiferromagnetism, and as a result the magnetic structure consists of ferromagnetically aligned spins, where the magnitudes of the ordered moments are sinusoidally modulated due to the presence of the antiferromagnetic component. Upon cooling below 11~K, the amplitude of this modulation decreases, and the system becomes closer to a uniform ferromagnet at low temperatures. This unusual temperature dependence may be a consequence of a change in the anisotropy of the exchange interactions, as also reflected by the variation of the incommensurate modulation with temperature. However, the anisotropy of the exchange interactions as well as the detailed characterization of the nature of the coexistence between ferromagnetism and  antiferromagnetism requires further study, in particular of single crystal samples.

We find throughout our study that the data are consistently well described with the crystal structure corresponding to the AlB$_2$-type structure, with a single inequivalent Nd site and no superlattice. This is consistent with a previous study of Nd$_2$PdSi$_3$ \cite{R2PdSi32}, although this compound has also been reported to have a $2\times2\times2$ superlattice structure \cite{R2PdSi36}. While evidence for such a superlattice with a doubling of all three axes has been reported for a number of $R_2TX_3$ materials \cite{Chevalier1984,R2PdSi36}, larger superlattices of  $2\times2\times4$ \cite{Gordon1997} or even $2\times2\times8$ have been reported for some other  $R_2$PdSi$_3$ compounds \cite{Tang}. In the case of  Nd$_2$PdSi$_3$, we note that the modulation of the ordered Nd moment cannot be accounted for by variations of the moment magnitude between different inequivalent Nd sites within a superstructure, since in this scenario the AFM component would be expected to have the same periodicity as the superstructure, but instead the antiferromagnetic propagation vector is incommensurate along the $c$-axis. On the other hand, to determine whether a superlattice is present in Nd$_2$PdSi$_3$, crystallographic studies of single crystals are very important, especially using local probes. In particular, it has been proposed that the superstructure may not be detectable in polycrystalline samples, due to shorter correlation lengths \cite{Tang}, and M\"{o}ssbauer effect measurements of Eu$_2$PdSi$_3$ revealed crystallographically inequivalent Eu sites, despite the corresponding superstructure Bragg peaks not being observed in powder x-ray diffraction \cite{Mallik1998c}.

The inelastic neutron scattering study shows the presence of CEF excitations in the paramagnetic state, with an additional excitation emerging below $T_0$. This data could be analyzed on the basis of a model for a Nd$^{3+}$ atom in a hexagonal CEF, allowing us to estimate the wave function of the ground state Kramer's doublet. The present study offers microscopic evidence that  Nd$_{2}$PdSi$_{3}$ exhibits complex magnetism, which is characterized by a dominant ferromagnetic contribution, as well as an antiferromagnetic component at the onset of  long range magnetic order, making it distinct from others in the $R_2$PdSi$_3$ family.

\begin{acknowledgments}
MS acknowledges funding support from the National Key R\&D Program of China (Grant No. 2017YFA0303100) and the  National Natural Science Foundation of China (Grant No. 11874320). DTA would like to thank  the Royal Society of London for the UK-China Newton mobility funding. DTA and ADH would like to thank CMPC-STFC, grant number CMPC-09108, for financial support. 

\end{acknowledgments}


\begin{thebibliography}{43}
\bibitem [{Ref()}]{RefNote1}%
  \BibitemOpen
  \href@noop {} {}\bibinfo {note} {Note in these high intensity data taken at
  25 K, there was also no sign of the superlattice peaks expected for $P63/mmc$
  symmetry with Pd and Si on separate Wyckoff sites.}\BibitemShut {Stop}%
\end{thebibliography}
\end{document}